\documentclass[prl,aps,superscriptaddress,twocolumn]{revtex4}
\usepackage{bm}
\usepackage{amsmath}
\usepackage{siunitx}
\usepackage{xcolor}
\usepackage{float}
\usepackage{graphicx}

\begin{document}
\title{How to unloop a self-adherent sheet}
\author{Twan J.S. Wilting}  
\affiliation{Fluids and Flows Group, Department of Applied Physics, J.M. Burgers Centre for Fluid Dynamics, Eindhoven University of Technology - Den Dolech 2, 5600 MB Eindhoven, The Netherlands.}
\author{Martin H. Essink} 
\affiliation{Physics of Fluids Group, Faculty of Science and Technology, J.M. Burgers Centre for Fluid Dynamics, Mesa+ Institute, University of Twente - P.O. Box 217, 7500 AE Enschede, The Netherlands.}
\author{Hanneke Gelderblom}  
\affiliation{Fluids and Flows Group, Department of Applied Physics, J.M. Burgers Centre for Fluid Dynamics, Eindhoven University of Technology - Den Dolech 2, 5600 MB Eindhoven, The Netherlands.}
\author{Jacco H. Snoeijer} \email{j.h.snoeijer@utwente.nl}
\affiliation{Physics of Fluids Group, Faculty of Science and Technology, J.M. Burgers Centre for Fluid Dynamics, Mesa+ Institute, University of Twente - P.O. Box 217, 7500 AE Enschede, The Netherlands.}
	
\date{\today}
	
\begin{abstract} 
The mechanics of adherent sheets is central to applications ranging from patching a band aid,  coating technology, to the breakthrough discovery of peeling graphene flakes using sticky tape. These processes are often hindered by the formation of blisters and loops, which are notoriously difficult to remove. Here we describe and explain a remarkable phenomenon that arises when one attempts to remove a loop in a self-adherent sheet that is formed by, e.g., folding two adhesive sides of a tape together. One would expect the loop to simply unloop when pulling on its free ends. Surprisingly, however, the loop does not immediately open up but shrinks in size, held together by a tenuous contact region that propagates along the tape. This adhesive contact region only ruptures once the loop is reduced to a critical size. We experimentally show that the loop-shrinkage results from an interaction between the peeling front and the loop, across the contact zone. This new type of interaction falls outside the realm of the classical elastica theory and is responsible for a highly nonlinear increase in the  peeling force. Our results reveal and quantify the increased force required to remove loops in self-adherent media, which is of importance for blister removal and  exfoliation of graphene sheets.
\end{abstract}

\maketitle

Folding, self-adhering, blistering, and peeling phenomena occur in many types of thin elastic layers, such as capillary films \cite{roman2010elasto, mora2012shape, bico2018elastocapillarity}, soft adhesives \cite{Dalbe_2015,creton2016fracture}, protective coatings or multi-layered materials \cite{schmidt2001thin,wagner_sticky_2013,davidovitcharxiv2020}, thin films floating on liquid or polymer substrates \cite{Pocivavsek, vellaPNAS}, or graphene sheets \cite{Li2009, Schniepp2008}. 
The mechanical properties and stability of these layers are crucial to applications such as thin flexible electronic devices \cite{plaut_postbuckling_2004, vellaPNAS}, the self-assembly of graphene ribbons \cite{annett2016self} or liquid-phase exfoliation of layered two-dimensional nanomaterials \cite{Botto}.  These processes are often hindered by delamination and the formation of blisters that are difficult to remove. In liquid-phase exfoliation, for example, graphene sheets can self-attach or reattach, and lead to the formation of unwanted self-adherent loops. 

\begin{figure}[ht]
    \includegraphics[page=1]{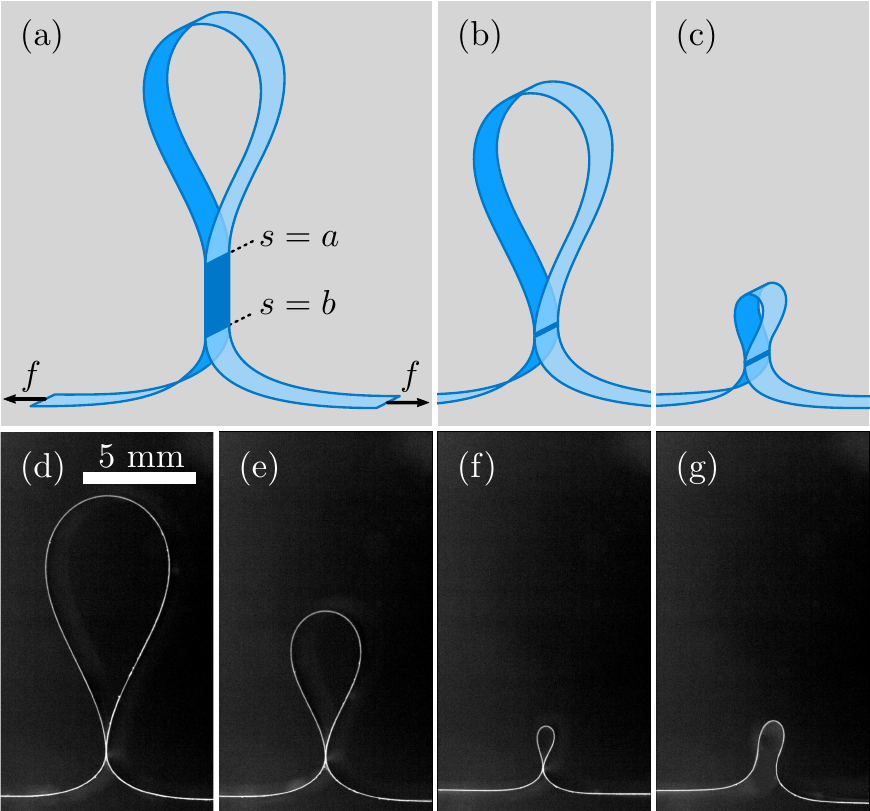}
    \caption{
    A sticky loop created by bonding two sides of an adhesive tape is difficult unloop. (a) When pulling the two ends with a force $f$, the adhesive contact zone decreases in length (dark blue, between $s=a$ and $s=b$), while the size of the loop initially remains constant. (b-c) When the contact zone becomes very small the loop does not unloop: instead, the loop shrinks in size, held together by a narrow adhesive contact that propagates along the tape. 
    (d-g) Experimental images of the shrinking loop and its eventual rupture. Only when the loop is reduced down to a critical size, it finally opens up. The images are captured for ``tape A" at a peeling velocity $v=4.2 \cdot 10^{-3}$~mm/s (cf. SI~Movie S1).}
    \label{fig:fig1}
\end{figure}

Here we describe and quantify a remarkable phenomenon that occurs when removing blisters by peeling. Consider an adhesive tape that is bent such that two sticky sides bond together, forming a loop, as shown in Fig.~\ref{fig:fig1}a. If one tries to open the loop by pulling the two loop ends apart, one encounters a rather unexpected dynamics. Initially, the extended adhesive zone where the two sides of the tape are in contact, decreases in size. At the moment where one would expect the loop to open up (Fig.~\ref{fig:fig1}b), unlooping does not occur: instead, the loop shrinks in size, held together by a narrow contact zone that propagates along the tape (Fig.~\ref{fig:fig1}c-f). This shrinking process continues until the loop reaches a critical size at which the contact eventually breaks, leading to the belated unlooping of the sticky tape (Fig.~\ref{fig:fig1}g). In spite of its importance for the applications mentioned above, and its ease to be reproduced, the peeling physics of such a self-adhered loop has so far received little attention \cite{bottega_peeling_1991}; for example, it is not known what is the force required to remove a self-adherent loop. Here we reveal the mechanism by which the contact zone of a loop of adhesive tape propagates and the loop shrinks, and we determine when the loop eventually breaks. 

\section{Experimental procedure} 
To create loops of adhesive tape we carefully fold the tape such that two sides are aligned and stick together, with a large extended region of self-contact, where the two ``contact lines" are well-separated. These contact lines are denoted by their arc-length coordinates, respectively, at $s=a$ and $b$ (insets of  Figs.~\ref{fig:imgs_overlay}~and~\ref{fig:fig5}). We then perform peeling experiments where two linear motors are used to pull both tape ends apart with a constant velocity $v$ (see SI~Movie~S.3). 
We characterize the evolution of the process by measuring the loop size and the curvatures on both sides of the contact zone. These curvatures are key quantities of the problem: they are known to offer a direct access to the peeling force whenever the tape -- outside the contact region -- is accurately described by the elastica ~\cite{Obreimoff_1930,wagner_sticky_2013,Dalbe_2015,perrin2019peeling}. The curvatures are therefore extracted from the experimental profiles through fits to the elastica equation. 
We extract the edge of the outer (non-adhesive) side of the tape from these images, and find the shape of the midplane by correcting for the thickness of the tape. The shape of the tape can indeed accurately be fitted as an elastica, and allow us to measure the loop size $a$ and the curvatures at the edge of the contact region $\kappa_{a,b}$. The curvature at the peeling front $s=b$ is determined on both sides of the symmetry plane and average values are reported (values that differ by more than 10\% are discarded). 

To investigate the robustness of the phenomenon we applied a broad range of peeling velocities ($v=4.2\cdot 10^{-4}$ to $1.2$~mm/s), and used three different types of commercial tapes that we indicate by tape A, B, C. Tape D is a thicker tape, formed by sticking two layers of tape A together (details given in Table \ref{tab:table3}). 
Most of the quantitative results will be presented for tape A, which has a thickness $t=\SI{46.0}{\micro\meter}$ and a width of $\SI{15.3}{\milli\meter}$ (similar results for other types are reported in the Supplement). In addition, we verified that contact zone propagation and delayed rupture also occur when peeling a loop of a non-adhesive elastic sheet that is held together by a lubricant (see SI, Movie~S.4). Due to an accumulation of lubricant in the contact zone during peeling, these experiments are only used for to demonstrate the universality of the phenomena but not for a quantitative analysis. The quantitative experiments are complemented by a theory that extends the classical elastica with a model for the contact region, to explain why the loop shrinks, when it ruptures, and what is the force required to unloop the loop. 

\begin{table}[h!]\centering
\caption{Technical specifications of the tapes}\label{tab:table3}
\begin{center}
\begin{tabular}{lccr}
Tape & Thickness  & Width & Brand \\ 
 & [$\SI{}{\micro\meter}$] & [$\SI{}{\milli\meter}$] & \\ 
\hline
A & 46.0  $\pm$ 0.5 & 15.3 $\pm$ 0.1 &  DLP Industry\\    
B & 43.0  $\pm$ 0.5 & 11.8 $\pm$ 0.1 &  Quantore\\  
C & 45.0  $\pm$ 0.5 & 15.2 $\pm$ 0.1 &  Tesa Film Basic\\ 
D & 92.0  $\pm$ 1.0 & 15.3 $\pm$ 0.1 &  DLP Industry (2$\times$)\\
\hline
\end{tabular}
\end{center}
\end{table}

\section{Shrinking dynamics and unlooping condition}Each experiment starts with an extended contact zone (dark blue region in Fig.~\ref{fig:fig1}a). 
When the applied peeling force is sufficiently large, the free end of tape peels at $s=b$ and the contact zone shortens (SI, Movie~S.3). During this initial stage the loop size, quantified by $s=a$, remains constant. However, as the contact zone becomes sufficiently small the two contact lines start to interact. The peeling front at $s=b$ induces a ``rolling" motion~\cite{bottega_peeling_1991} during which the contact line position  $s=a$ is also being displaced. This effective interaction between the two contact lines typically starts when the size of the contact falls below ten times the tape thickness. At this moment, the large loop begins to decrease in size and follows the sequence shown in Fig.~\ref{fig:fig1}. During this process, the size of the tenuous adhesive zone gradually decreases, until it finally ruptures (cf. SI Movie~S.2).

\begin{figure}
	\centering
    \includegraphics[page=2]{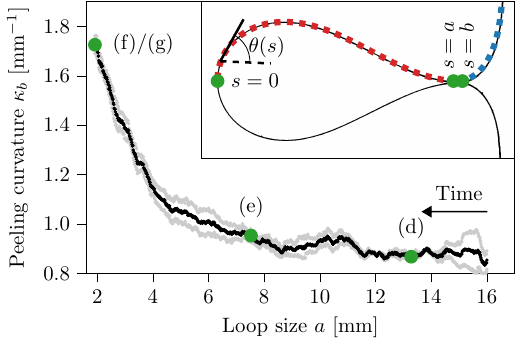}
	\caption{
		Peeling curvature $\kappa_b$
		as a function of the loop size $a$. Green dots correspond to the snapshots in Fig.~\ref{fig:fig1}, from which the data in this figure is obtained. The curvature at the peeling front increases as the loop shrinks, which indicates a strong increase of the peeling force. The average peeling curvature is shown in black, with the curvature in 
		$s=b$ on both sides of the symmetry plane shown in gray. Inset: Experimental tape profile fitted by the elastica equation (red and blue curves). The size of the loop is quantified by the arc-length coordinate of the contact line at $s=a$, while the peeling front is at $s=b$. The corresponding curvatures $\kappa_a$ and $\kappa_b$, respectively, are determined from the elastica fits.
	}
	\label{fig:imgs_overlay}
\end{figure}

Surprisingly, we find that during the experiment the curvature at the peeling front ($s=b$) increases, which implies a strongly nonlinear increase of the peeling force. This increase can be seen upon careful inspection of Fig.~\ref{fig:fig1}, and is further quantified in Fig.~\ref{fig:imgs_overlay}, where we plot the curvature $\kappa_b$ at the peeling front as a function of the loop size $a$. Initially, the curvature remains approximately constant, but it steeply increases prior to the unlooping. Such an increase is in contrast with a previous analysis of the loop mechanics \cite{bottega_peeling_1991}. There, the curvature at $s=b$ was predicted to remain constant, namely $\kappa_b=\sqrt{2\gamma/B}$, with $\gamma$ the adhesion energy per unit area (or fracture energy necessary for debonding), and $B$ the bending modulus of the tape. 
The same expression for the curvature was found for blisters \cite{wagner_sticky_2013}, peeling~\cite{Dalbe_2015} and elastocapillary loops \cite{roman2010elasto,bico2018elastocapillarity}, and actually goes back as far as the mechanics of splitting mica \cite{Obreimoff_1930}. Importantly, each of these situations involve only a single, \emph{isolated} peeling front with an infinitely extended contact (i.e.~$s=b$ without a second contact at $s=a$). In the present context, the isolated peeling front curvature $\kappa_{\rm iso}=\sqrt{2\gamma/B}$ is obtained in the limit where the distance between the two contact lines is still large. This limit is approached at large loop sizes, so that $\kappa_{\rm iso}$ corresponds to the large-$a$ plateau in Fig.~\ref{fig:imgs_overlay}. Importantly, the increase of curvature~--~robustly observed for all tapes (cf. SI~Fig.~S.1a)~--~implies a strong interaction between the two contact lines at $s=a$ and $b$, which remains to be explained.

\begin{figure}
	\centering
    \includegraphics[page=3]{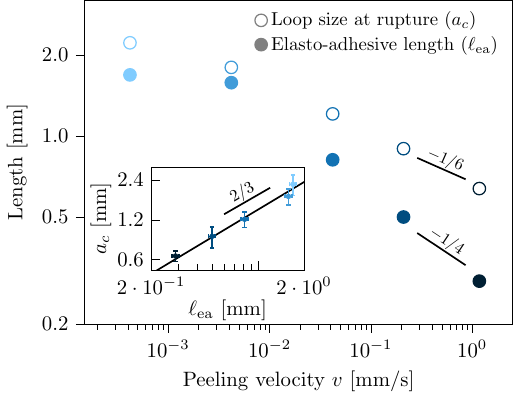}
	\caption{
		Effect of the peeling velocity $v$ on the critical loop size at rupture $a_c$ (open symbols) and the elasto-adhesive length $\ell_{\rm ea}$ (defined in text, closed symbols). Both quantities decrease with $v$ (slopes are a guide to the eye). Inset: Relation between $a_c$ and $\ell_{\rm ea}$. A power-law fit gives an exponent $0.70\pm 0.09$, consistent with the exponent 2/3 that is predicted by the model \eqref{eq:scaling_ac}. All data  for tape A and each datapoint represents an average over at least 10 experiments.
	}
	\label{fig:fig3}
\end{figure}

The unlooping process exhibits an intricate dependence on the peeling velocity. The open symbols in Fig.~\ref{fig:fig3} show the critical loop size $a_c$, taken at the point of rupture, as a function of the peeling velocity $v$. The results are from the same tape as in Figs.~\ref{fig:fig1} \& \ref{fig:imgs_overlay}, and each datapoint represents an average over at least 10 experiments. Clearly, the critical loop size depends on the peeling velocity: faster peeling enables smaller loops. We attribute this trend to an increase of adhesion energy $\gamma$ with velocity. The adhesion energy for pressure-sensitive adhesives generically exhibits a power-law dependence with peeling velocity, which originates from the strong dissipation that occurs during debonding when polymers are pulled out of the adhesive matrix~\cite{Maugis_1978,Gent_1996,creton2016fracture}. A stronger dissipative adhesion makes it more difficult to break the contact, leading to smaller loops. To verify this hypothesis, we determine the \emph{elasto-adhesive length} $\ell_{\rm ea}\equiv \sqrt{B/\gamma}$ of the tape from the curvature of an isolated peeling front, so that we can use $\kappa_{\rm iso} = \sqrt{2}/\ell_{\rm ea}$~\cite{bottega_peeling_1991,wagner_sticky_2013,Dalbe_2015,roman2010elasto}. Specifically, we measure $\ell_{\rm ea}$ from the large-$a$ plateau in Fig.~\ref{fig:imgs_overlay}. The closed symbols in Fig.~\ref{fig:fig3} show that $\ell_{\rm ea}$ decreases with $v$, in a way that is consistent with a typical dissipation $\gamma \sim v^{0.5}$~\cite{Maugis_1978,Gent_1996,creton2016fracture,perrin2019peeling}. Unexpectedly, however, the loop size at rupture is not simply proportional to the elasto-adhesive length. As is shown in the inset of Fig.~\ref{fig:fig3}, the best power-law fit gives $a_c \sim \ell_{\rm ea}^{0.7}$. This power-law dependence implies that, besides $\ell_{\rm ea}$, another length scale must be involved, and points to physics beyond the classical elastica theory.

\begin{figure}
	\centering
	\includegraphics[page=4]{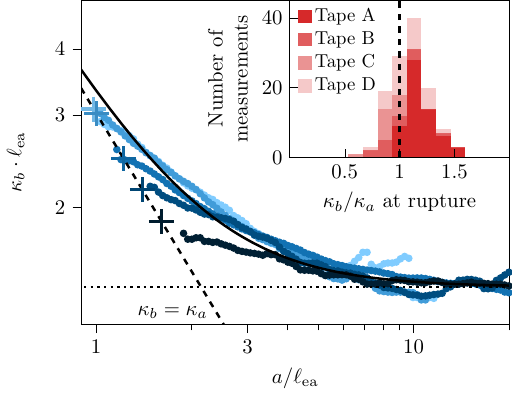}
	\caption{
		Dimensionless plot of peeling curvature $\kappa_b \ell_{\rm ea}$ versus loop size $a/\ell_{\rm ea}$. The data correspond to different peeling velocities (same color code as in Fig.~\ref{fig:fig3}, all for tape A). For each dataset, the smallest value of $a$ indicates the point of rupture, which occurs when $\kappa_b$ is close to $\kappa_a=3.028/a$ (dashed line). The symbols ``+" indicate the model prediction for rupture (with $\alpha=4.3$, see text). The solid line is the simplified model \eqref{eq:naive}, the dotted horizontal line shows $\kappa_{\rm iso}=\sqrt{2}/\ell_{\rm ea}$. Inset: Histogram of $\kappa_b / \kappa_a$ at rupture for various $v$ and for the four tapes.
	}
	\label{fig:fig4}
\end{figure}

Having established the velocity-dependence of the elasto-adhesive length $\ell_{\rm ea}$, we try to collapse the peeling dynamics. Fig.~\ref{fig:fig4} reports the curvature $\kappa_b$ versus the loop-size $a$ for different $v$, non-dimensionalized by  $\ell_{\rm ea}$. The datasets do not collapse; we see a systematic trend as the imposed peeling velocity is increased (from light to dark blue). However, the experiments do reveal a striking common feature: rupture always occurs close to the point where $\kappa_b = \kappa_a$, where $\kappa_a\sim 1/a$ is indicated by the dashed line. This observation suggests that  \textbf{unlooping occurs when the curvatures are approximately equal} on both sides of the contact. While there is some variability between individual experiments, there is strong evidence for this equal-curvature hypothesis. Figure~\ref{fig:fig4} (inset) reports the histogram over 123 experimental realizations, obtained for the four different tapes. The histogram peaks near $\kappa_a=\kappa_b$, with a small bias to unloop slightly before reaching the point of equal curvatures.

\section{Macroscopic model} 
Now we explain these robust  observations using a mechanical model that describes the macroscopic (outer) scale of the loop problem. Specifically, we aim to describe to find an explanation for the increase in peeling curvature as the loop shrinks, the critical size of the loop, and, correspondingly, the critical peeling force.

The bending energy $\mathcal E_B$ of the tape and the work $\mathcal W$ performed by the external peeling force $f$ (all quantities taken per unit tape width), read
\begin{equation}\label{eq:EB}
\mathcal E_B = \int_0^L ds \, \frac{1}{2}B\theta'^2, \quad \quad 
\mathcal W = \int_0^L ds \, f \sin \theta.
\end{equation}
Here the angle $\theta(s)$ gives the shape of the tape (see inset Fig.~\ref{fig:imgs_overlay}), $B$ is the tape's bending modulus, while $s=L$ is the end of the tape where the peeling force is applied. In the regions where the tape does not stick to itself, the shape follows from the minimisation of the total mechanical energy with respect to $\theta(s)$, i.e. $\delta\left( \mathcal E_B - \mathcal W\right)/\delta \theta=0$. This minimisation gives the classical elastica equation~\cite{audoly2010elasticity}
\begin{equation}\label{eq:elastica}
B \theta'' + f \cos \theta =0.
\end{equation}
Solutions to this equation indeed provide excellent fits of the tape~\cite{wagner_sticky_2013}, as is clear from the dashed lines in the inset of Fig.~\ref{fig:imgs_overlay}\footnote{Since the loop is not directly connected to the peeling force, due to the contact zone, \eqref{eq:elastica} for $s<a$ requires a different value of $f$, which turns out to be $f=9.91 B/a^2$.}. Importantly,  \eqref{eq:elastica} can be integrated once to $f=\frac{1}{2}B \theta'(b)^2=\frac{1}{2}B \kappa_b^2$ \cite{Dalbe_2015}. Hence, the increase of $\kappa_b$ is indeed a direct measurement of the increase in the peeling force. 

We now turn to the peeling itself. At $s=b$ peeling amounts to a displacement of the contact line by a distance $-d b$, taken as a positive quantity, during which mechanical energy is released. When the peeling is overdamped, so that inertia plays no role, the release of mechanical energy ${G_b =  \partial (\mathcal E_B - \mathcal W)/\partial b}$ is exactly equal to the adhesion energy  $\gamma(v)$. This analysis is in direct analogy to fracture mechanics, where the quantity $G_b$ is referred to as the energy release rate associated to the propagation of a crack tip~\cite{eshelby1975elastic,rice1968path}. The energy \eqref{eq:EB} gives (derivation in SI),
\begin{equation}\label{eq:Gb}
G_b = \frac{\partial (\mathcal E_B - \mathcal W)}{\partial b} = \frac{1}{2}B \kappa_b^2.
\end{equation}
Equating $G_b=\gamma$, one indeed recovers the isolated contact line condition $\kappa_{\rm iso} = \sqrt{2}/\ell_{\rm ea}$~ \cite{wagner_sticky_2013,Dalbe_2015}. 
To include the motion of the contact line at $s=a$, we consider the ``rebonding" that occurs when the two sides of the tape are pushed together. Debonding and rebonding are asymmetric processes: debonding  occurs when polymers are pulled out of a matrix and it typically leads to fibril formation, while the rebonding is comparatively gentle and involves a negligible amount of dissipation. This difference in dissipation is responsible for the difference in curvature at the advancing and receding sides of the peeling front. In analogy to \eqref{eq:Gb}, one finds  the energy release rate during rebonding at ${s=a}$ to be
\begin{equation}\label{eq:Ga}
G_a = \frac{\partial (\mathcal E_B - \mathcal W)}{\partial a} = -\frac{1}{2}B \kappa_a^2.
\end{equation}
The appearance of a minus sign implies that bending energy is actually being \emph{stored} rather than released, as the loop shrinks. Since the adhesive energy gained by rebonding is negligible, this storage of elastic energy must originate from an interaction with contact line at $b$, which pushes the contact line at $a$. Crucially, however, this interaction is not accounted for in \eqref{eq:EB}, so that additional physics is needed to explain the shrinking of the loop.

As a simple model, we first neglect the finite size of the contact region. We assume that the loop starts to shrink once $b=a$, and that  subsequently these points are displaced together. The corresponding propagation condition, $G_b+G_a=\gamma$, gives 
\begin{equation}\label{eq:naive}
\frac{1}{2}B(\kappa_b^2 - \kappa_a^2)= \gamma,
\end{equation}
and is shown as the solid line in Fig.~\ref{fig:fig4}. This result already offers an excellent description of the data at large loop-sizes. Specifically, (\ref{eq:naive}) explains the increase of peeling curvature (and thus of the peeling force): this increase can be attributed to the storage of elastic energy inside the shrinking loop. 

However, \eqref{eq:naive} does not allow for equal curvatures $\kappa_b=\kappa_a$, and does not predict any rupture of the loop. To refine the analysis one needs to account for the physics inside the contact zone, where one encounters complexities associated to the finite thickness of the tape, the viscoelastic shearing and stretching of the adhesive, and the extraction of polymers during debonding \cite{Maugis_1978,Gent_1996,creton2016fracture,perrin2019peeling,kaelble1960theory, villey2015rate, villey2017situ, ghatak2003adhesion, adda2006crack}. 
Here we explore the feasibility of a generic outer description of the problem that allows to explain the robust macroscopic observations without being specific on the scale of the adhesive -- just like the outer scale of the intricate dynamical debonding of a single front is captured by a single macroscopic length $\ell_{\rm ea}$. 
To this end we propose an effective macroscopic interaction energy, $\mathcal E_{\rm int}$, that  accounts for the interaction between the contact lines at $a$ and $b$. With this energy, the propagation conditions become
\begin{alignat}{4}
& \frac{\partial (\mathcal E_B +\mathcal E_{\rm int} - \mathcal W)}{\partial b} &&= 
 &&\frac{1}{2}B \kappa_b^2 + \frac{\partial \mathcal E_{\rm int}}{\partial b} &&= \gamma \label{eq:b}, \\
& \frac{\partial (\mathcal E_B +\mathcal E_{\rm int} - \mathcal W)}{\partial a} &&= 
 - &&\frac{1}{2}B\kappa_a^2 + \frac{\partial \mathcal E_{\rm int}}{\partial a} &&=0. \label{eq:a}
\end{alignat}
When the interaction energy is only a function of the distance $w=b-a$, one recovers \eqref{eq:naive} and no progress is made. A key observation is that in experiments the length of the contact $w$ reaches a scale comparable to that of thickness $t$ of the tape and the adhesive, so that the system can no longer be described as an infinitely thin elastica. As a phenomenological closure, we therefore hypothesize that any difference in curvature $\kappa_b-\kappa_a$ over such a short distance $w \sim t$ comes with an extra elastic energy beyond \eqref{eq:EB}. In the spirit of a gradient expansion, we account for this \emph{gradient} in curvature by an extra energy $\sim \int_a^b ds \, (\theta'')^2$, from which, using that $\theta'' \sim (\kappa_b-\kappa_a)/w$, we obtain the estimate 
\begin{equation} \label{eq:non-naive}
\mathcal E_{\rm int} = \alpha^2 t^2 B \frac{(\kappa_b-\kappa_a)^2}{b-a}.
\end{equation}
Here $t^2B$ is introduced on dimensional grounds, so that $\alpha$ is a dimensionless constant. The interaction \eqref{eq:non-naive} expresses that a difference in curvature cannot be sustained for $w \ll t$, and allows for rupture only when $\kappa_b =\kappa_a$.  

The proposed $\mathcal E_{\rm int}$ gives an accurate description of the experiments (derivations in SI). First, far from rupture, the model reduces to (\ref{eq:naive}) and thus explains the collapse in Fig.~\ref{fig:fig4} at large loop size (as also observed for tapes  B, C, and D, cf. Fig.~S.1 in the Supplement). Second, the model describes the loop rupture, with critical values of the loop size $a_c$ and $\kappa_b$ as indicated by the ``+" symbols in Fig.~\ref{fig:fig4}, with $\alpha=4.3$ as a single adjustable parameter (we take $t$ as the tape thickness). Hence, the model captures the intricate velocity dependence of the experiment, as well as the equal-curvature condition. Third, for  $t/\ell_{ea} \ll 1$ the model reduces to a scaling law (cf. SI), 
\begin{equation} \label{eq:scaling_ac}
     a_c \sim t^{1/3}\ell_{\rm ea}^{2/3},
\end{equation}
which is in close agreement with the experimental results as shown for tape A in the inset of Fig.~\ref{fig:fig3}. As such, the proposed gradient expansion for $\mathcal E_{\rm int}$ captures the essential macroscopic features of the shrinking and subsequent unlooping.

\begin{figure}
	\centering
	\includegraphics[page=5]{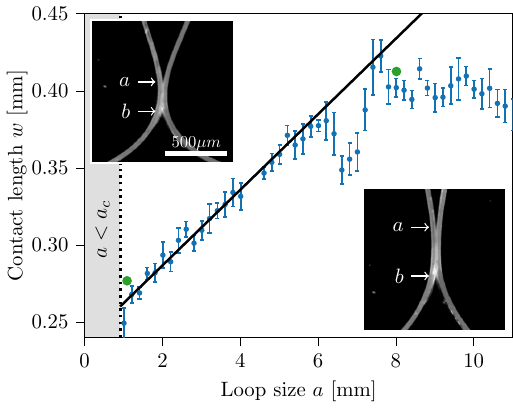}
	\caption{
	    Contact length $w$ as a function of loop size $a$, for tape A at a velocity of $v=4.2 \cdot 10^{-2}$~mm/s. The contact size decreases linearly upon approaching the point of rupture $a_c$ (solid line is the best fit). The data is binned over intervals in $a$, with errorbars indicating the 95\% confidence interval. The insets provide typical snapshots near the contact region, with arrows marking the estimated locations of the contact lines. The corresponding data points of the insets are indicated in green in the main figure.
	}
	\label{fig:fig5}
\end{figure}

To achieve a more detailed understanding of physics inside the contact zone, one could complement the current results with experiments and theory on the inner problem, on the scale of the adhesive, as previously achieved for single peeling fronts  \cite{kaelble1960theory, villey2015rate, villey2017situ, ghatak2003adhesion, adda2006crack, creton2016fracture}. As a first step in this direction, we here explore the contact length $w=b-a$, which in the model is predicted to vanish at the point of rupture, according to

\begin{equation} \label{eq:scaling_w}
    w \sim \left(\frac{t}{\ell_{\rm ea}}\right)^{4/3}(a-a_c).
\end{equation}
To verify this prediction of the contact length experimentally, one needs to bear in mind that in the macroscopic description the points $s=a,b$ formally represent the extrapolation of the outer elastica profiles (see inset Fig.~\ref{fig:imgs_overlay}). In practice, however, the small difference between the two extrapolated values $s=a,b$ turns out too sensitive to give reproducible results. As an alternative, we therefore determine the length of the contact from close-up recordings near the contact region. The insets of Fig.~\ref{fig:fig5} illustrate that the positions of the two ``fronts" are not sharply defined, due to the finite thickness of the tape and the deformed adhesive; see e.g.~\cite{villey2015rate, villey2017situ, ghatak2003adhesion} for a more detailed view on a single peeling front. Here we determined the position of the contact lines by taking the outside location of the bright region, as indicated by the arrows. The resulting contact length $w$ is reported in  Fig.~\ref{fig:fig5}. The contact length indeed decreases linearly close to the point of rupture, but does not seem to vanish completely at the critical loop size. This offset could possibly be attributed to a systematic overestimation induced by the local measurement of the contact positions. For example, an uncertainty of 20 $\mu$m in the direction of the thickness of the tape/adhesive, would already lead to an error of $\sqrt{20 \, \mu{\rm m}/\kappa_b} \sim 100 \, \mu{\rm m}$ on the position of each of the contact lines. Future work on the detailed contact mechanics could shed further light on these  microscopic features. The macroscopic experiments presented here, and the outer description in terms of a gradient expansion, will serve as clear benchmarks.

\section{Conclusion}
In summary, we have analyzed the shrinkage and subsequent unlooping of a self-adhered elastic tape. We have shown that the phenomenon of shrinking is mediated by a tenuous contact zone, whose mechanics is not part of the classical elastica theory. The model proposed for the contact zone offers a good description of the experimental observations, and explains the critical size and critical force at which the tape unloops. Importantly, our findings are not restricted to a fold of sticky tape. The exact same effect of contact propagation arises in non-adhesive loops held together by a lubricant (see SI Movie S.4) or when an adhesive sheet does not adhere to itself, but to another surface -- as one verifies using sticky tape on a table and exerting a force on each end of the tape (see SI Movie S.5). Therefore, we expect that our findings will be applicable to a broader range of problems involving loops and blisters. These phenomena appear in e.g.~coatings, flexible electronics and during the exfoliation of graphene sheets \cite{hernandez2008high}, for which narrow adhesive zones plays a crucial role in the appearance of folded and scrolled loop topologies \cite{li2018mechanics,yi2014temperature}.

\acknowledgments
T.J.S.W. contributed equally to this work with M.H.E. The authors gratefully acknowledge discussions with L. Botto, A. Darhuber and A. Pandey. H.G. acknowledges financial support from the Netherlands Organisation for Scientific Research (NWO), Veni Grant No. 680-47-451. M.H.E. and J.H.S. acknowledge financial support from ERC (the European Research Council) Consolidator Grant No. 616918, and from NWO through Vici Grant No. 680-47-632.
 
\bibliographystyle{apsrev4-1}
\bibliography{UnloopStickyTape}

\end{document}